\documentclass[reprint, aps, prd, showpacs, superscriptaddress, floatfix]{revtex4-2}

\usepackage{graphicx}
\usepackage{amsmath}
\usepackage{amssymb}

\usepackage[caption=false]{subfig}
\usepackage{subcaption}
\usepackage[colorlinks=true, citecolor=blue, urlcolor=blue, linkcolor=blue]{hyperref}

\begin{document}

\title{Detector Dependence of Inspiral Christodoulou Gravitational Wave Memory in Binary Black Hole Systems}

\author{Jiswin Varghese}
\email{jiswinvarghese@gmail.com}
\affiliation{Department of Physics, Government College, Kattappana, Idukki, Kerala 685508, India}

\author{Jyothipriya M. Shaji}
\email{jyothipriya276@gmail.com}

\author{Lyjo K Joseph}
\email{lyjo@gckottayam.ac.in}
\affiliation{Department of Physics, Government College, Kattappana,Idukki, Kerala 685508, India }

\date{\today}

\begin{abstract}










The nonlinear gravitational-wave memory, also known as the Christodoulou memory effect, is a permanent displacement produced by the self-interaction of gravitational waves predicted by General Relativity. Although its theoretical properties have been extensively investigated, the dependence of the observable inspiral memory on binary parameters and detector sensitivity remains an active area of study. In this work, we present GWMemoryLab, a modular numerical framework developed to investigate the leading-order Christodoulou memory generated during the inspiral phase of non-spinning binary black hole systems within the post-Newtonian approximation.
The framework implements physically motivated modules describing the binary dynamics, post-Newtonian inspiral evolution, oscillatory waveform generation, gravitational-wave energy flux, and nonlinear memory accumulation. The numerical implementation is validated through analytical comparisons and convergence tests, demonstrating excellent numerical stability. We investigate the dependence of the accumulated memory on binary mass ratio, total mass, and detector low-frequency cutoff. The simulations reveal the existence of an optimal total mass that maximizes the observable inspiral memory for a given detector bandwidth. Furthermore, the numerical results indicate that the optimal mass is well described by an approximately inverse dependence on the detector low-frequency cutoff,  $M_{\rm peak} \propto f_0^{-1}$, within the parameter space explored. This behaviour is interpreted as the consequence of the competition between increasing gravitational-wave luminosity and decreasing inspiral duration as the binary approaches the innermost stable circular orbit (ISCO). These results provide physical insight into the detectability of nonlinear gravitational-wave memory and demonstrate the usefulness of modular numerical frameworks for systematic parameter studies.
\end{abstract}

\maketitle 

\section{Introduction}

The direct detection of gravitational waves has opened an unprecedented observational window into the strong-field regime of General Relativity. Since the first detection of gravitational waves from the binary black hole merger GW150914 by the LIGO Scientific and Virgo Collaborations, numerous compact binary coalescences have been observed, providing stringent tests of gravitational physics while establishing gravitational-wave astronomy as a powerful tool for exploring the Universe \cite{Abbott2016Observation, Abbott2019Catalog} .

One of the most intriguing nonlinear predictions of General Relativity is the gravitational-wave memory effect. Unlike the oscillatory component of a gravitational wave, which produces a transient deformation of spacetime, the memory effect produces a permanent displacement of freely falling test particles after the passage of the wave. This phenomenon arises from the nonlinear nature of Einstein's field equations and represents one of the most distinctive signatures of gravitational-wave self-interaction.

The nonlinear memory, commonly known as the Christodoulou memory, was first derived by Christodoulou, who demonstrated that gravitational waves themselves contribute to the stress-energy responsible for generating an additional non-oscillatory component of the spacetime metric \cite{Christodoulou1991}. Subsequent theoretical investigations established the connection between nonlinear memory and the total gravitational-wave energy radiated by compact binary systems, leading to detailed analytical descriptions within the post-Newtonian framework \cite{Blanchet1992,Thorne1992,Favata2009b, Favata2011}.

Although considerable theoretical progress has been achieved, numerical investigations that systematically examine the dependence of the observable inspiral memory on binary parameters and detector characteristics remain comparatively limited. In particular, the influence of the detector low-frequency cutoff on the accumulated nonlinear memory has received relatively little attention despite its importance for present and future ground-based gravitational-wave observatories.

In this work, we develop a modular numerical framework, GWMemoryLab, \cite{Varghese2026GWMemoryLab},  to investigate the leading-order Christodoulou memory generated during the inspiral phase of non-spinning binary black hole systems. The framework combines post-Newtonian orbital evolution, gravitational-wave energy flux calculations, oscillatory waveform generation, and nonlinear memory accumulation within a unified computational architecture. Extensive validation against analytical predictions and numerical convergence studies demonstrates the reliability of the implementation.

Using this framework, we perform a systematic exploration of the dependence of the inspiral memory on the binary mass ratio, total mass, and detector low-frequency cutoff. The simulations reveal an optimal total mass that maximizes the observable inspiral memory for a fixed detector bandwidth. Furthermore, we identify an approximately inverse scaling between the optimal total mass and the detector low-frequency cutoff over the parameter space investigated, providing a physical interpretation based on the competing effects of increasing gravitational-wave luminosity and decreasing inspiral duration.

The remainder of this paper is organized as follows. Section 2 presents the theoretical framework underlying the post-Newtonian inspiral and nonlinear memory calculations. Section 3 describes the numerical implementation of GWMemoryLab and the validation procedures. Section 4 presents the numerical results together with their physical interpretation. Finally, Section 5 summarizes the principal conclusions and discusses possible extensions of the present work.

\section{Theoretical Framework}

In this work, the gravitational-wave memory is computed within the leading-order post-Newtonian approximation for non-spinning binary black hole systems in quasi-circular orbits. The inspiral dynamics are modeled using the quadrupole approximation, while the nonlinear (Christodoulou) memory is obtained from the cumulative gravitational-wave energy radiated during the inspiral phase.

The present framework is restricted to the inspiral regime and therefore does not include the merger and ringdown phases. Consequently, the computed memory corresponds to the inspiral contribution to the nonlinear gravitational-wave memory.

\subsection{Binary System}

Consider a binary consisting of two compact objects with component masses $m_1$ and $m_2$. The total mass of the system is

\begin{equation}
    M=m_1+m_2,
\end{equation}

while the symmetric mass ratio is

\begin{equation}
    \eta=\frac{m_1m_2}{M^2},
\end{equation}

which satisfies

\begin{equation}
    0<\eta\leq\frac14.
\end{equation}

The chirp mass,

\begin{equation}
    \mathcal{M}=M\eta^{3/5},
\end{equation}

plays a central role in determining the inspiral evolution since it governs the leading-order frequency evolution of the binary \cite{Blanchet2014}.

\subsection{Post-Newtonian Inspiral}

The orbital evolution is described by the leading-order post-Newtonian frequency evolution equation \cite{Peters1963,Blanchet2014}. 

\begin{equation}
\frac{df}{dt}=\frac{96}{5}\pi^{8/3}\left(\frac{G\mathcal{M}}{c^3}\right)^{5/3}f^{11/3}
\label{eq:dfdt}
\end{equation}

where $f$ denotes the gravitational-wave frequency.

Equation (\ref{eq:dfdt}) is integrated numerically using a fourth-order Runge--Kutta scheme until the binary reaches the Schwarzschild innermost stable circular orbit (ISCO), whose gravitational-wave frequency is given by \cite{Misner1973, Maggiore2007}

\begin{equation}
    f_{\rm ISCO} = \frac{c^3}{6^{3/2}\pi GM}.
\end{equation}

The ISCO frequency defines the termination of the inspiral phase within the present model.

\subsection{Oscillatory Gravitational-Wave Signal}

The leading-order gravitational-wave strain is follows the quadrupole approximation, with the amplitude scaling as $f^{2/3}$ and the phase obtained from the time integral of the instantaneous frequency \cite{Maggiore2007,Blanchet2014}.

\begin{equation}
    h(t)=A(t)\cos[\phi(t)],
\end{equation}

where the phase evolves according to

\begin{equation}
    \phi(t) = 2\pi \int f(t)\,dt
\end{equation}

and the waveform amplitude follows the quadrupole scaling

\begin{equation}
    A(t)\propto f^{2/3}.
\end{equation}

These relations reproduce the characteristic chirp behaviour of compact binary inspirals.

\subsection{Gravitational-Wave Energy Flux}

The leading-order gravitational-wave luminosity is given by the quadrupole formula \cite{Peters1963,Blanchet2014}.

\begin{equation}
    \frac{dE}{dt}=\frac{32}{5}\frac{c^5}{G}\eta^2x^5,
\end{equation}
where
\begin{equation}
    x=\left(\frac{\pi GMf}{c^3}\right)^{2/3}
\end{equation}
is the standard post-Newtonian expansion parameter.
The cumulative radiated energy is obtained through numerical integrating the gravitational-wave luminosity over the inspiral evolution \cite{Blanchet2014, Maggiore2007}.

\begin{equation}
    E_{\rm GW}(t)=\int_0^t\frac{dE}{dt'}dt'
\end{equation}

\subsection{Christodoulou Memory}

The nonlinear (Christodoulou) memory arises from the energy carried by gravitational waves and produces a permanent change in the spacetime metric \cite{Christodoulou1991,Blanchet1992,Thorne1992,Favata2011,Favata2009b}.

\begin{equation}
    h_{\rm mem}=\frac{4GE_{\rm GW}}{c^4D},
    \label{eq:memory}
\end{equation}

where $D$ denotes the luminosity distance to the source.

Equation (\ref{eq:memory}) provides a direct relationship between the accumulated radiated energy and the observable nonlinear memory strain. Consequently, any parameter affecting the radiated energy also influences the final memory amplitude. It represents an angle-averaged / characteristic scalar scaling model for the cumulative memory strain amplitude. In full tensorial perturbation theory, gravitational-wave memory appears as a directional tensor perturbation decomposed into spin-weighted spherical harmonics ${}_{-2}Y_{lm}(\iota, \phi)$, with the dominant contribution arising from the $(l=2, m=0)$ mode. Treating Eq.~(\ref{eq:memory}) as a phenomenological characteristic scale allows us to isolate the dominant mass-energy dependence without loss of generality for isotropic comparisons.

\section{Numerical Method}

To investigate the inspiral contribution to the Christodoulou gravitational-wave memory, we developed a modular numerical framework named \texttt{GWMemoryLab}. The framework was designed to provide a transparent and extensible implementation of the leading-order post-Newtonian formalism while allowing systematic investigations of the dependence of the nonlinear memory on binary parameters and detector characteristics.

Each physical process is implemented as an independent module, enabling verification of individual components and facilitating future extensions to higher-order post-Newtonian models or more sophisticated waveform approximations.

\subsection{Software Architecture of GWMemoryLab}
\begin{figure}[t]
\centering
\includegraphics[width=1\linewidth]{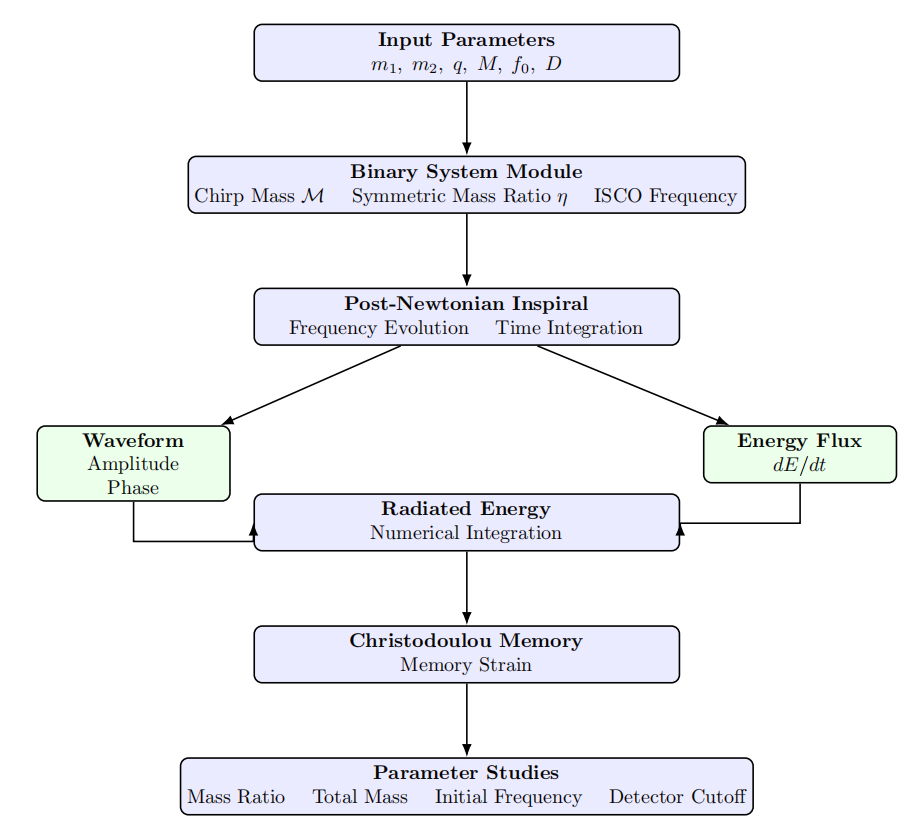}
\caption{
Architecture of the \texttt{GWMemoryLab} framework. Beginning with the physical parameters of the binary system, the framework computes the post-Newtonian inspiral evolution, generates the oscillatory gravitational-wave signal, evaluates the gravitational-wave energy flux, accumulates the emitted gravitational-wave energy, and finally determines the inspiral contribution to the Christodoulou gravitational-wave memory. The resulting framework enables systematic parameter studies of binary mass ratio, total mass, and detector low-frequency sensitivity.
}
\label{fig:architecture}
\end{figure}

The computational workflow adopted in the present study is illustrated in Fig.~\ref{fig:architecture}. Starting from the physical parameters of the binary system, the framework computes the post-Newtonian inspiral evolution, constructs the oscillatory gravitational-wave signal, evaluates the gravitational-wave energy flux, and determines the cumulative radiated energy. The accumulated Christodoulou memory is then obtained from the total radiated energy. Finally, the framework performs systematic parameter studies to investigate the dependence of the memory on the binary mass ratio, total mass, and detector low-frequency cutoff.

The modular design allows each physical stage of the calculation to be independently validated while maintaining a clear correspondence between the numerical implementation and the underlying theoretical formulation.

\subsection{Numerical Implementation}

The numerical framework was developed in Python using an object-oriented and modular architecture. Each physical component of the calculation is implemented as an independent module, allowing individual verification of the underlying equations while facilitating future extensions to higher-order post-Newtonian models.

The principal modules of \texttt{GWMemoryLab} are summarized below.

\begin{itemize}

\item \textbf{BinarySystem}: Computes the physical properties of the binary, including the total mass, symmetric mass ratio, chirp mass, and Schwarzschild ISCO frequency.

\item \textbf{Inspiral}: Evolves the gravitational-wave frequency using the leading-order post-Newtonian evolution equation until the binary reaches the innermost stable circular orbit.

\item \textbf{Waveform}: Generates the oscillatory gravitational-wave strain from the instantaneous frequency evolution by computing both the waveform amplitude and phase.

\item \textbf{EnergyFlux}: Evaluates the gravitational-wave luminosity throughout the inspiral and numerically integrates the emitted energy.

\item \textbf{Memory}: Computes the inspiral contribution to the nonlinear (Christodoulou) gravitational-wave memory from the accumulated radiated energy.
\end{itemize}
The modular structure allows each physical stage of the calculation to be independently tested while maintaining a direct correspondence between the numerical implementation and the theoretical formulation presented in Section~2.

\subsection{Verification and Numerical Accuracy}

To ensure the reliability of the numerical implementation, several verification tests were performed before the parameter studies.

First, the post-Newtonian inspiral evolution was verified by confirming that the gravitational-wave frequency increases monotonically throughout the inspiral and terminates at the Schwarzschild ISCO frequency predicted by General Relativity.

The oscillatory waveform exhibits the expected chirp behaviour, with both the frequency and amplitude increasing as the binary approaches coalescence.

The gravitational-wave luminosity increases continuously during the inspiral, while the cumulative radiated energy grows monotonically, demonstrating the consistency of the numerical integration.

The nonlinear memory strain also increases monotonically with time, reflecting the cumulative nature of the Christodoulou memory effect.

Finally, the numerical convergence of the inspiral integration was investigated by repeating the calculations using several integration time steps. The relative error decreases systematically as the time step is reduced, demonstrating stable numerical convergence.

\begin{table}[t]
\centering
\caption{Convergence study for the inspiral integration. The reference solution corresponds to the smallest integration time step.}
\begin{tabular}{cccc}
\hline
Time Step (s) &
Memory &
Steps &
Relative Error \\
\hline

$10^{-3}$ &
$3.466885\times10^{-22}$ &
926 &
$5.015\times10^{-5}$ \\

$5\times10^{-4}$ &
$3.466755\times10^{-22}$ &
1851 &
$1.245\times10^{-5}$ \\

$10^{-4}$ &
$3.466713\times10^{-22}$ &
9251 &
$3.771\times10^{-7}$ \\

$5\times10^{-5}$ &
$3.466711\times10^{-22}$ &
18501 &
Reference \\

\hline
\end{tabular}
\label{tab:convergence}
\end{table}

The convergence results demonstrate that the numerical solution is insensitive to the integration time step for sufficiently small values of $\Delta t$. Throughout the parameter studies presented in this work, an integration time step of $10^{-4}\,\mathrm{s}$ was adopted, providing an excellent compromise between computational efficiency and numerical accuracy.

\section{Results and Discussion}

The numerical framework described in the previous sections was used to investigate the inspiral contribution to the Christodoulou gravitational-wave memory for non-spinning binary black hole systems. Unless otherwise stated, the calculations assume quasi-circular binaries located at a luminosity distance of 400 Mpc and employ an integration time step of $10^{-4}$ s. The inspiral evolution is terminated at the Schwarzschild innermost stable circular orbit (ISCO).

\subsection{Post-Newtonian Inspiral Evolution}
\begin{figure}
    \centering
    \includegraphics[width=0.9\linewidth]{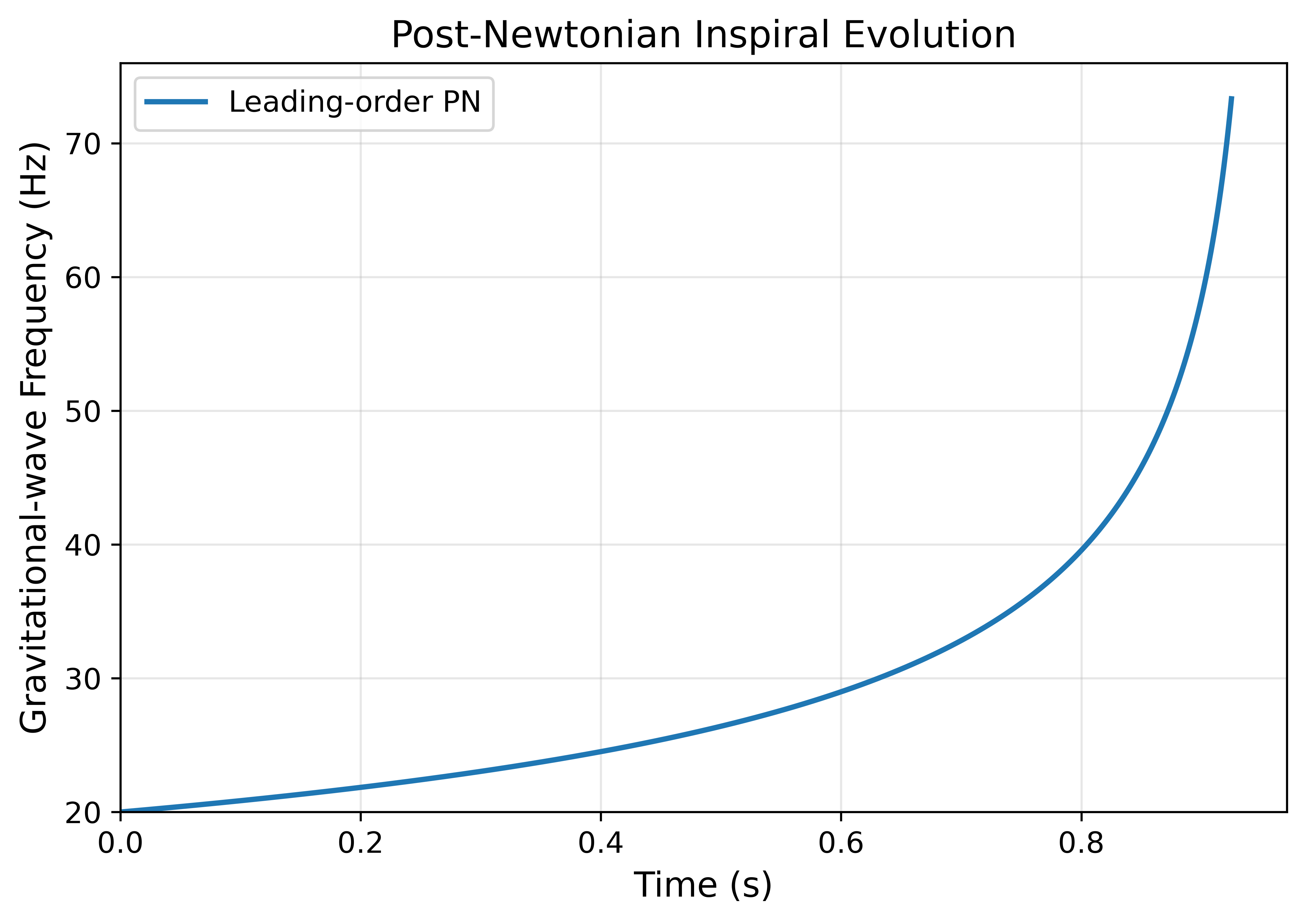}
    \caption{Evolution of the gravitational-wave frequency during the inspiral phase for an equal-mass binary black hole system. The frequency increases monotonically owing to gravitational-wave energy loss and terminates at the Schwarzschild innermost stable circular orbit (ISCO), marking the end of the inspiral approximation. }
    \label{fig:pn_evolution}
\end{figure}

Figure~\ref{fig:pn_evolution} shows the gravitational-wave frequency as a function of time for an equal-mass binary black hole system. The frequency increases continuously throughout the inspiral, exhibiting the characteristic chirp behaviour predicted by post-Newtonian theory. As the binary loses orbital energy through gravitational-wave emission, the orbital separation decreases, causing the orbital and gravitational-wave frequencies to increase progressively.
The numerical evolution terminates at the Schwarzschild ISCO, which defines the limit of validity of the inspiral approximation adopted in the present work. The monotonic increase of the frequency confirms the stability of the numerical integration and reproduces the expected leading-order post-Newtonian dynamics.

The inspiral evolution provides the fundamental input for all subsequent calculations, including the waveform generation, gravitational-wave luminosity, cumulative radiated energy, and nonlinear memory accumulation.
The rapid increase in frequency near the end of the inspiral is consistent with the leading-order post-Newtonian prediction,
\[
\frac{df}{dt}\propto f^{11/3},
\]
which implies an accelerating inspiral as the binary approaches coalescence.

\subsection{Oscillatory Gravitational-Wave Signal}

The leading-order inspiral waveform follows the quadrupole approximation for compact binary systems \cite{Maggiore2007,Blanchet2014}.

\begin{figure}
    \centering
    \includegraphics[width=0.9\linewidth]{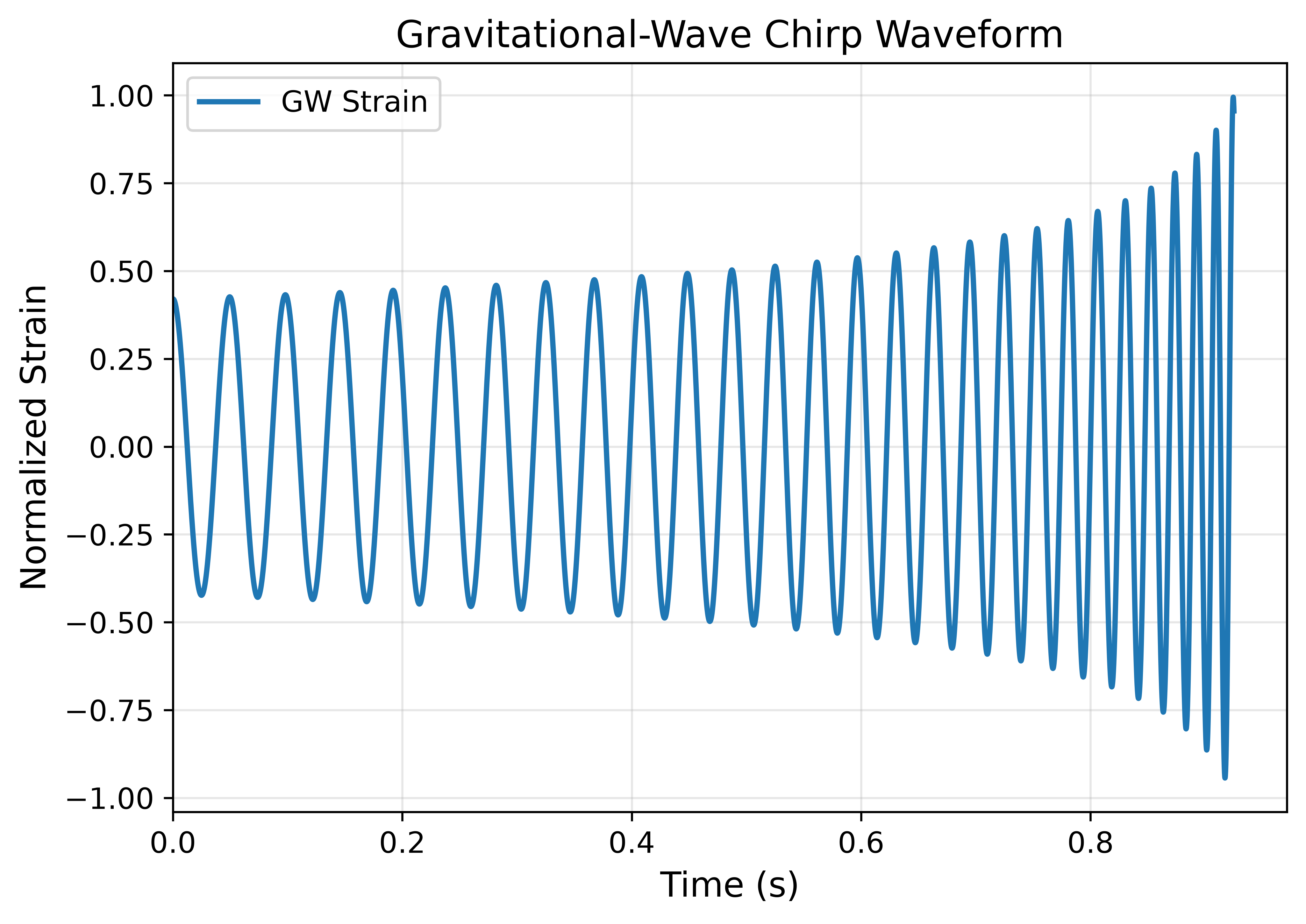}
    \caption{Oscillatory gravitational-wave strain generated from the leading-order post-Newtonian inspiral. The waveform exhibits the characteristic chirp behaviour of compact binary inspirals, with increasing frequency and amplitude as the binary approaches the innermost stable circular orbit.}
    \label{fig:waveform}
\end{figure}

The oscillatory gravitational-wave strain generated from the post-Newtonian inspiral is shown in Fig.~\ref{fig:waveform}. The waveform exhibits the characteristic chirp signature of compact binary inspirals, with both the oscillation frequency and amplitude increasing as the binary evolves toward the innermost stable circular orbit.

The increasing frequency arises from the gradual reduction of the orbital separation caused by gravitational-wave emission, while the growth in amplitude follows the leading-order quadrupole relation $h \propto f^{2/3}$. Consequently, the waveform becomes progressively compressed in time and larger in amplitude as the inspiral proceeds.

The successful reproduction of the expected chirp behaviour demonstrates that the inspiral evolution, phase integration, and waveform generation modules operate consistently within the numerical framework. Since the subsequent calculations of gravitational-wave luminosity and nonlinear memory are based on the same inspiral evolution, the waveform provides an additional validation of the physical implementation.

\subsection{Gravitational-Wave Energy Flux}
\begin{figure}
    \centering
    \includegraphics[width=0.9\linewidth]{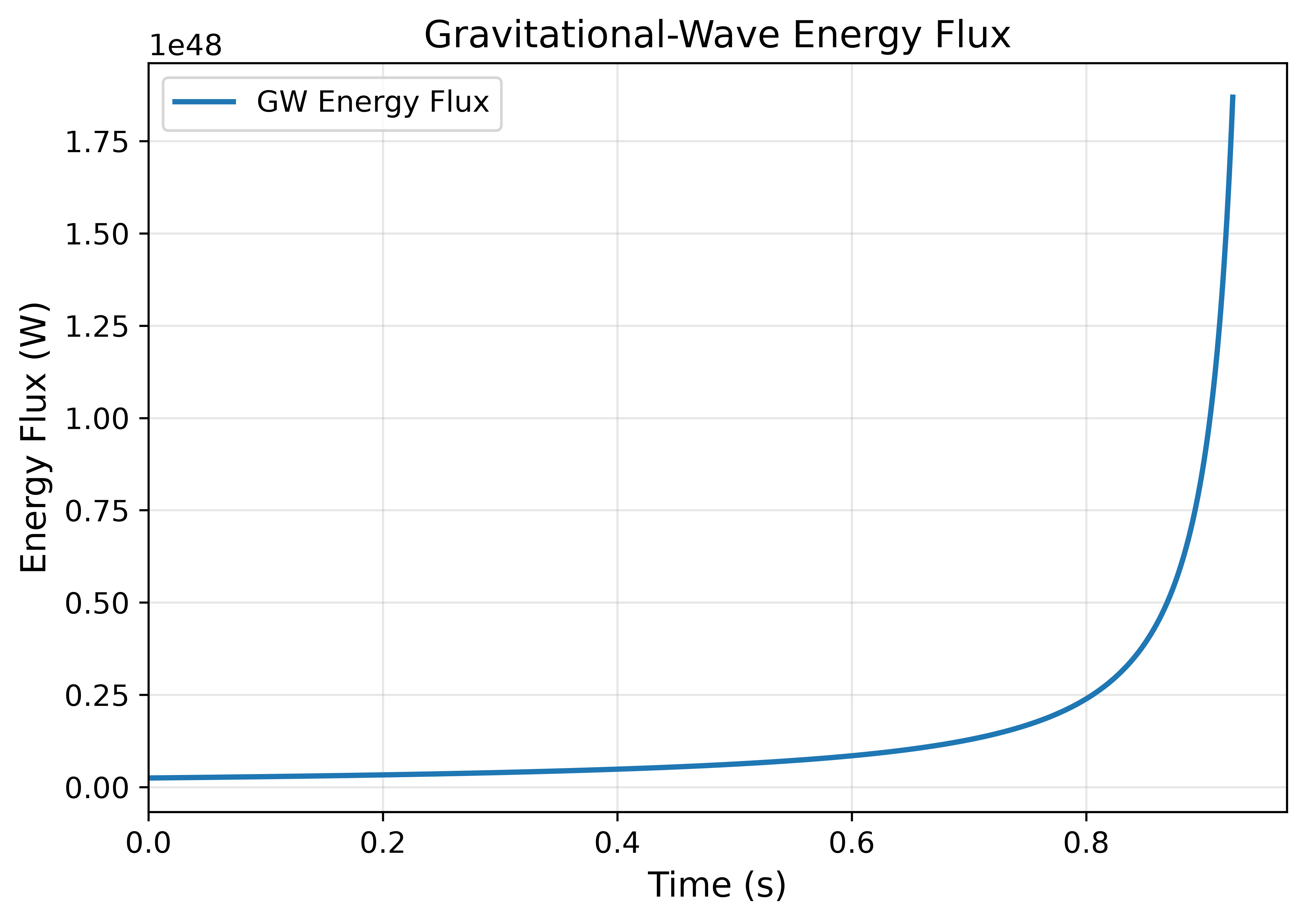}
    \caption{Instantaneous gravitational-wave energy flux during the inspiral phase. The emitted luminosity increases rapidly as the orbital frequency rises, reaching its maximum immediately before the inspiral terminates at the Schwarzschild ISCO.}
    \label{fig:energy_flux}
\end{figure}
Figure~\ref{fig:energy_flux} presents the instantaneous gravitational-wave energy flux during the inspiral evolution. The luminosity increases monotonically throughout the inspiral, reaching its maximum immediately before the binary reaches the Schwarzschild ISCO.

This behaviour is a direct consequence of the increasing orbital velocity as the binary components spiral inward. Since the leading-order gravitational-wave luminosity scales approximately as
\[
\frac{dE}{dt}\propto f^{10/3},
\]
the rapid increase in gravitational-wave frequency near the end of the inspiral produces a corresponding enhancement in the emitted power.

The numerical results reproduce the expected physical behaviour predicted by the quadrupole approximation. Because the Christodoulou memory depends on the cumulative energy transported by gravitational waves, the increasing luminosity shown in Fig.~\ref{fig:energy_flux} provides the physical mechanism responsible for the subsequent growth of the nonlinear memory.
The monotonic increase in the numerical luminosity agrees with the analytical prediction of the leading-order quadrupole formula, confirming that the numerical implementation correctly captures the expected inspiral dynamics.

\subsection{Cumulative Radiated Energy}

\begin{figure}
    \centering
    \includegraphics[width=0.9\linewidth]{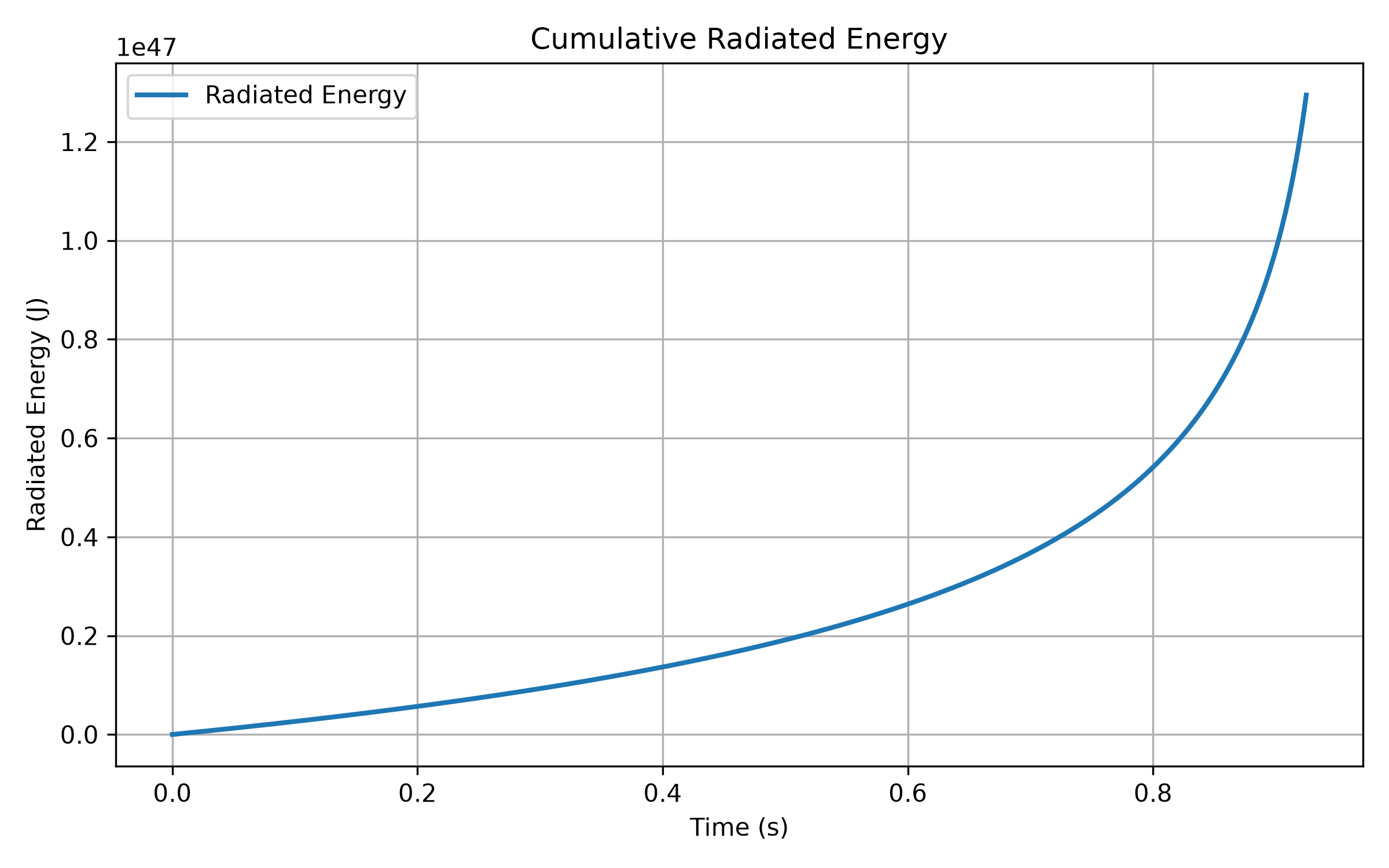}
    \caption{Cumulative gravitational-wave energy radiated during the inspiral phase. The emitted energy increases monotonically as gravitational-wave emission continuously extracts orbital energy from the binary system, reaching its maximum at the end of the inspiral.}
    \label{fig:radiated_energy}
\end{figure}

The cumulative gravitational-wave energy radiated during the inspiral is shown in Fig.~\ref{fig:radiated_energy}. Unlike the instantaneous luminosity, which represents the rate of energy emission, the cumulative radiated energy increases monotonically throughout the inspiral as gravitational-wave emission continuously extracts orbital energy from the binary system.

At early times, the radiated energy increases gradually because the orbital velocity and gravitational-wave luminosity are relatively small. As the binary approaches the innermost stable circular orbit, the inspiral accelerates and the luminosity rises rapidly, producing a steeper increase in the cumulative radiated energy. The final value therefore represents the total gravitational-wave energy emitted during the inspiral phase within the leading-order post-Newtonian approximation.

Since the nonlinear Christodoulou memory is directly related to the total energy carried away by gravitational waves, the cumulative radiated energy provides the physical quantity from which the memory strain is subsequently determined. Consequently, the behaviour shown in Fig.~\ref{fig:radiated_energy} establishes the foundation for understanding the monotonic growth of the nonlinear memory presented in the following subsection.

\subsection{Growth of the Christodoulou Memory}

\begin{figure}
    \centering
    \includegraphics[width=0.9\linewidth]{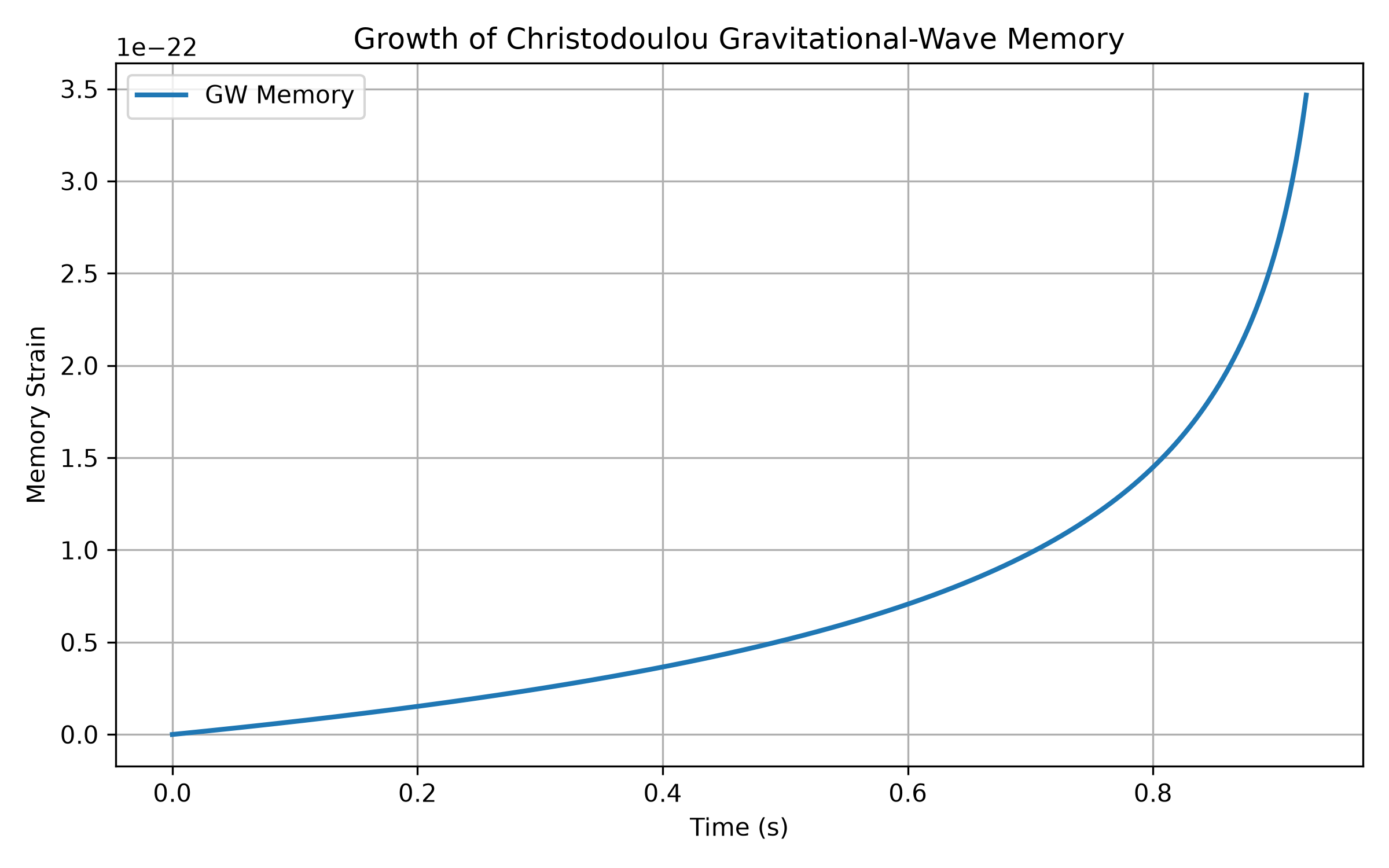}
    \caption{Evolution of the Christodoulou gravitational-wave memory during the inspiral phase. The memory strain increases monotonically because it is determined by the cumulative gravitational-wave energy emitted by the binary, approaching its final value at the end of the inspiral. }
    \label{fig:memory_growth}
\end{figure}

Figure~\ref{fig:memory_growth} presents the evolution of the nonlinear (Christodoulou) gravitational-wave memory during the inspiral. In contrast to the oscillatory gravitational-wave strain, which alternates between positive and negative values, the memory increases continuously throughout the inspiral and approaches a constant final value at the termination of the simulation.

The monotonic behaviour reflects the cumulative nature of the nonlinear memory effect. Since the memory is generated by the energy transported by gravitational waves, every increment in the emitted gravitational-wave energy contributes positively to the accumulated memory strain. As a result, the memory cannot decrease during the inspiral and instead grows steadily as additional energy is radiated.

The growth rate becomes progressively larger toward the end of the inspiral owing to the rapid increase in gravitational-wave luminosity. Consequently, the largest contribution to the final memory is produced during the late inspiral, where the binary radiates energy most efficiently before reaching the Schwarzschild ISCO.

The monotonic evolution observed in Fig.~\ref{fig:memory_growth} is consistent with the theoretical interpretation of the Christodoulou memory as a hereditary effect, whose value depends on the entire previous history of gravitational-wave emission rather than the instantaneous state of the binary. \cite{Christodoulou1991,Thorne1992,Blanchet1992,Favata2009b}

\subsection{Dependence on Binary Mass Ratio}

\begin{figure}
    \centering
    \includegraphics[width=0.9\linewidth]{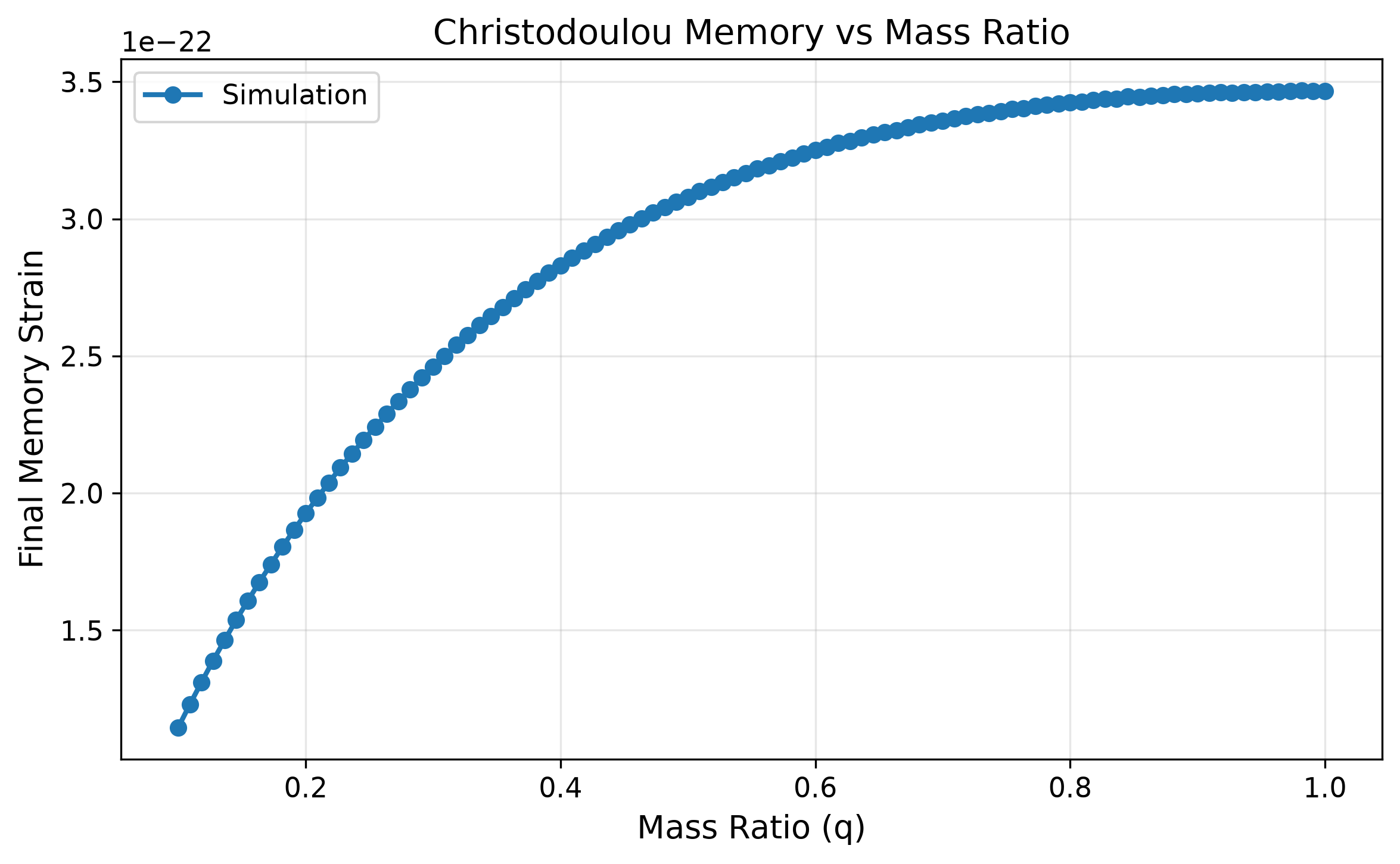}
    \caption{Final Christodoulou memory strain as a function of the binary mass ratio. The nonlinear memory increases monotonically with increasing mass ratio and reaches its maximum for equal-mass binaries (\$q=1\$), reflecting the enhanced gravitational-wave emission associated with the maximum symmetric mass ratio. }
    \label{fig:memory_mass_ratio}
\end{figure}

The dependence of the final Christodoulou memory on the binary mass ratio is shown in Fig.~\ref{fig:memory_mass_ratio}. The calculations were performed for binary systems with mass ratios ranging from $q=0.01$ to $q=1$, while maintaining a fixed total mass and detector configuration.

The numerical results show that the accumulated nonlinear memory increases monotonically with increasing mass ratio, reaching its maximum for the equal-mass configuration ($q=1$). In contrast, binaries with highly unequal masses generate significantly smaller memory amplitudes.

This behaviour can be understood from the dependence of the gravitational-wave emission on the symmetric mass ratio,
\begin{equation}
    \eta=\frac{q}{(1+q)^2},
\end{equation}
which reaches its maximum value of $\eta=0.25$ for equal-mass binaries. Since the gravitational-wave luminosity scales approximately as $\eta^2$ at leading post-Newtonian order, equal-mass systems radiate gravitational-wave energy more efficiently than unequal-mass binaries. Because the Christodoulou memory is generated by the cumulative radiated energy, binaries with larger symmetric mass ratios naturally produce stronger nonlinear memory signals.

The numerical results therefore confirm the theoretical expectation that equal-mass binaries constitute the most efficient inspiral sources of Christodoulou memory within the parameter space investigated. This trend is consistent with previous theoretical studies of nonlinear gravitational-wave memory from compact binary systems.

The numerical data are well described by a simple empirical relation between the final memory strain and the binary mass ratio. The fitted model reproduces the numerical results with an excellent coefficient of determination ($R^2 \approx 0.99999$), indicating that the memory varies smoothly over the investigated parameter range. This agreement demonstrates the internal consistency of the numerical framework and provides a convenient parameterization of the inspiral memory for phenomenological applications.
\cite{Blanchet2014,Favata2009b,Maggiore2007}

\subsection{Dependence on the Initial Frequency}

\begin{figure}
    \centering
    \includegraphics[width=0.9\linewidth]{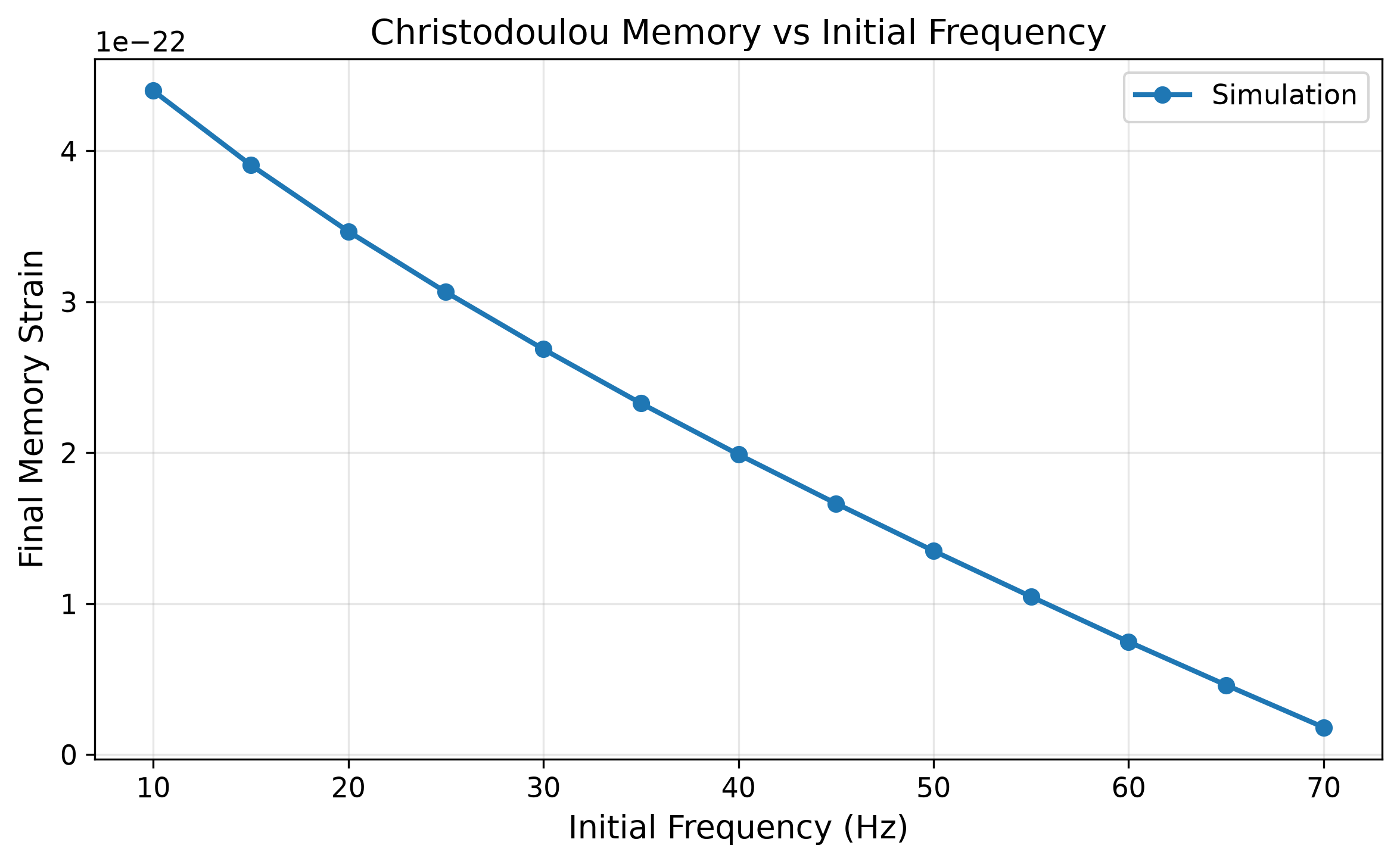}
    \caption{Final Christodoulou memory strain as a function of the initial gravitational-wave frequency. The accumulated memory decreases monotonically as the initial frequency increases because a progressively smaller fraction of the inspiral contributes to the observed gravitational-wave energy. }
    \label{fig:memory_frequency}
\end{figure}

The influence of the initial gravitational-wave frequency on the accumulated Christodoulou memory is shown in Fig.~\ref{fig:memory_frequency}. The calculations were performed for a fixed binary configuration while varying the initial frequency from 10 Hz to 70 Hz. In the context of gravitational-wave observations, the initial frequency corresponds to the detector's low-frequency sensitivity limit, since the inspiral evolution is only observed once the signal enters the detector bandwidth.

The numerical results demonstrate a clear monotonic decrease in the accumulated memory as the initial frequency increases. For an initial frequency of 10 Hz, the inspiral memory reaches approximately $4.40\times10^{-22}$, whereas for an initial frequency of 70 Hz the memory is reduced to approximately $1.80\times10^{-23}$, corresponding to more than an order of magnitude decrease.

This behaviour is expected from the cumulative nature of the Christodoulou memory. Lower initial frequencies allow the detector to observe a larger fraction of the inspiral, thereby including a longer period of gravitational-wave emission and a greater accumulation of radiated energy. Conversely, increasing the initial frequency excludes the early inspiral evolution from the calculation, reducing the total radiated energy contained within the observed signal and consequently decreasing the final memory strain.

The results emphasize that the observable inspiral memory depends not only on the intrinsic properties of the binary but also on the frequency band accessible to the detector. Consequently, detectors with improved low-frequency sensitivity are expected to recover a significantly larger fraction of the nonlinear gravitational-wave memory.

These results demonstrate that the detector bandwidth plays a fundamental role in determining the observable inspiral memory. Since the nonlinear memory is a hereditary effect that accumulates throughout the inspiral, extending the observational band toward lower frequencies enables a larger fraction of the accumulated memory to be measured. This dependence motivates the systematic investigation of the combined influence of detector cutoff and binary mass presented in the following subsection. \cite{Favata2011,Blanchet2014,Maggiore2007}

\subsection{Total Mass Scaling of the Inspiral Memory}

\begin{figure}
    \centering
    \includegraphics[width=0.9\linewidth]{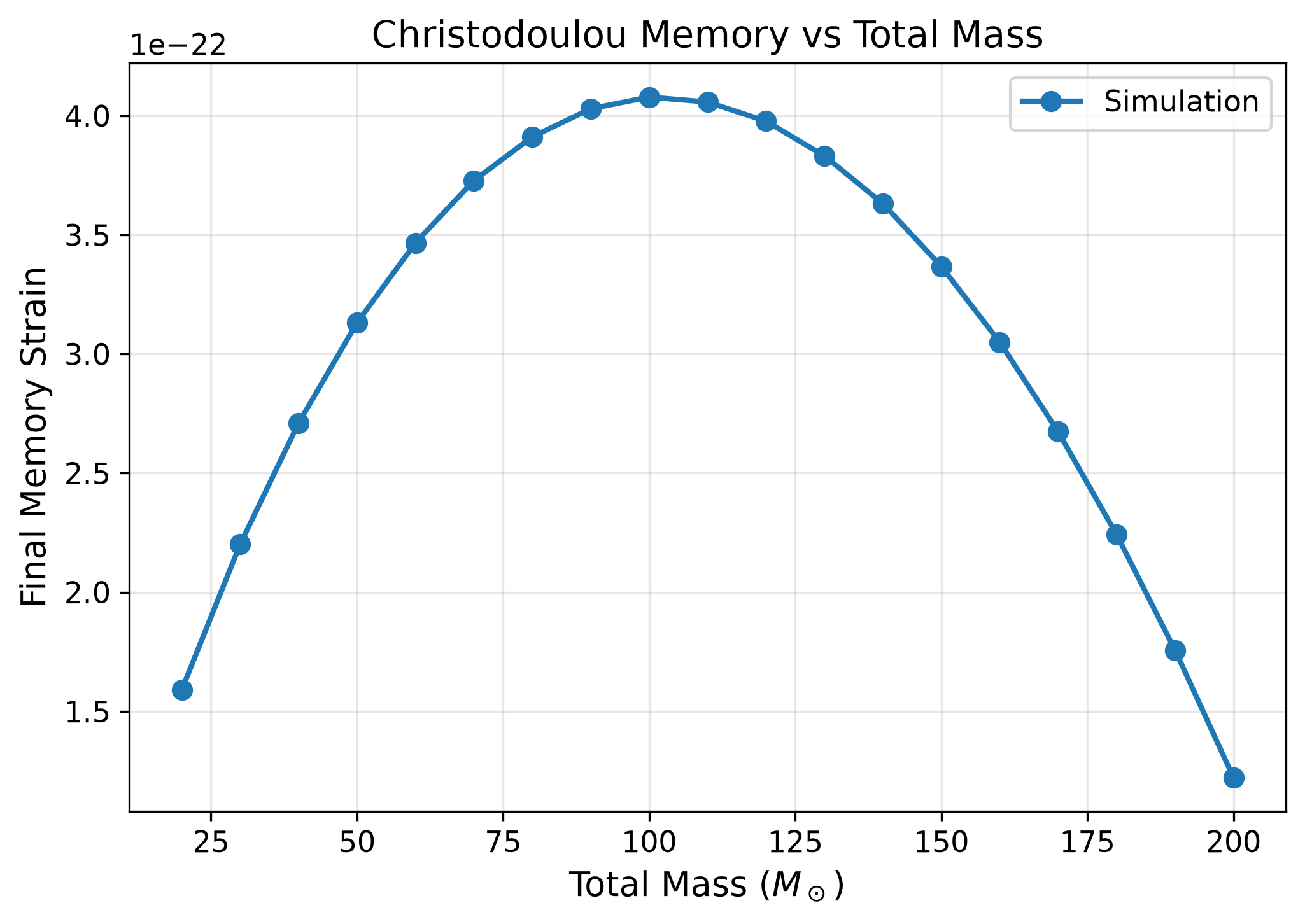}
    \caption{Final Christodoulou memory strain as a function of the total binary mass for an initial frequency of 20 Hz. The memory increases with mass at lower masses, reaches a maximum near $100M\odot$, and subsequently decreases because the inspiral duration becomes progressively shorter as the ISCO frequency approaches the detector's low-frequency cutoff. }
    \label{fig:memory_total_mass}
\end{figure}

The dependence of the final Christodoulou memory on the total binary mass is presented in Fig.~\ref{fig:memory_total_mass}. The calculations were performed for equal-mass binary black holes with total masses ranging from $20\,M_{\odot}$ to $200\,M_{\odot}$ while maintaining a fixed initial frequency of 20 Hz.

\begin{figure}
    \centering
    \includegraphics[width=1\linewidth]{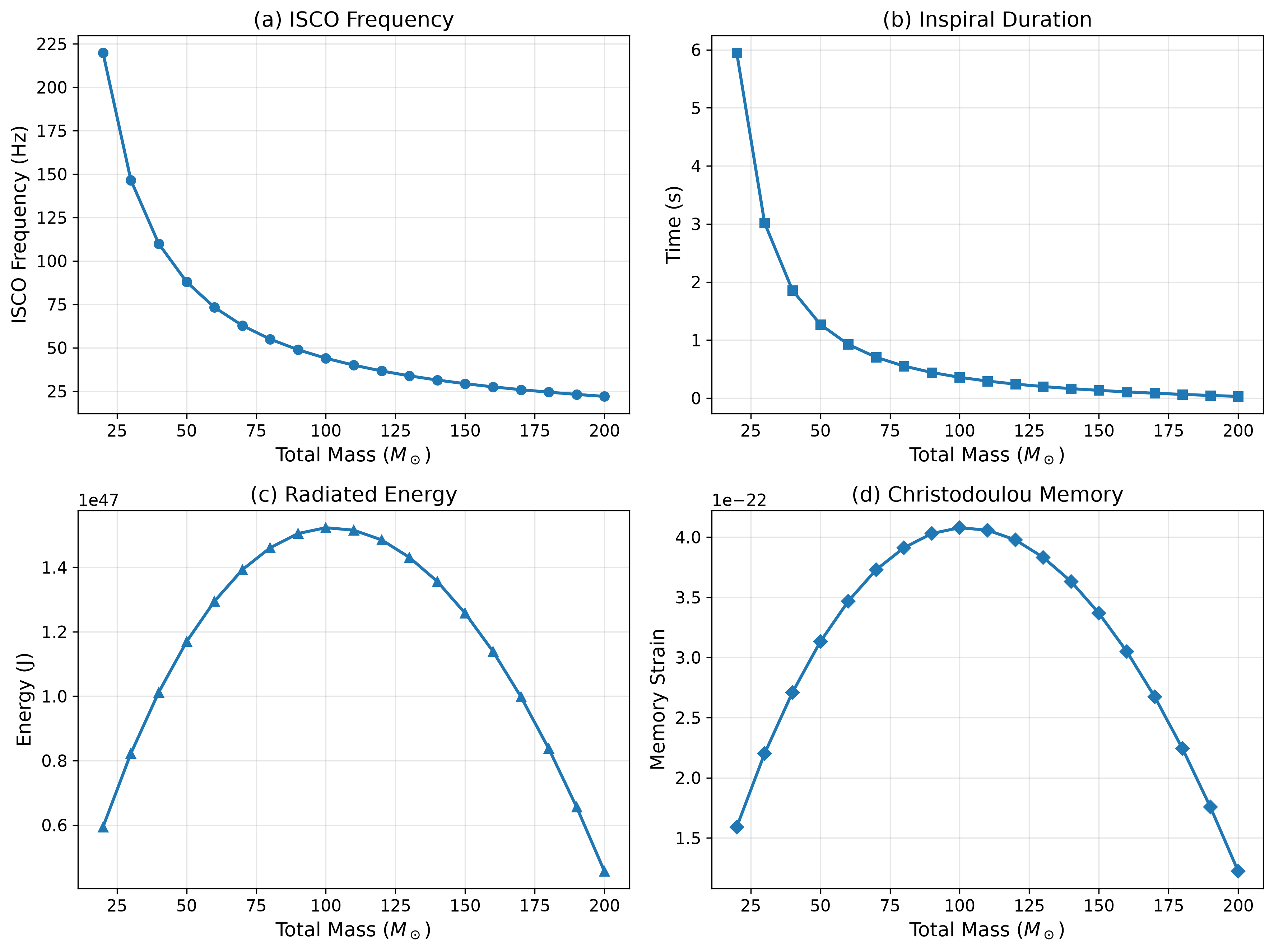}
    \caption{
            Diagnostic breakdown of binary inspiral properties as a function of total binary mass (M) for an initial detector frequency of $f\_0 = 20\text{ Hz}$. (a) Innermost stable circular orbit (ISCO) frequency ($f_{\text{ISCO}}$), showing a monotonic $\propto M^{-1}$ decay. (b) Total inspiral duration within the detector bandwidth, rapidly decreasing for higher masses. (c) Total cumulative radiated gravitational-wave energy ($E_{GW}$), exhibiting a peak at intermediate mass. (d) Final Christodoulou memory strain ($h_{\text{mem}}$), directly tracking the radiated energy profile and reaching a maximum near $M = 100M\odot$.}
    \label{fig:mass_scaling}
\end{figure}
The numerical results reveal a non-monotonic dependence of the inspiral memory on the total mass. As the total mass increases from $20\,M_{\odot}$, the memory strain initially grows and reaches a maximum value of approximately $4.08\times10^{-22}$ near a total mass of $100\,M_{\odot}$. For higher masses, however, the accumulated memory decreases despite the larger gravitational-wave luminosity associated with more massive binaries.

This behaviour results from the competition between two opposing physical effects. Increasing the total mass enhances the gravitational-wave emission, tending to increase the accumulated memory. At the same time, the Schwarzschild ISCO frequency decreases according to

\begin{equation}
f_{\rm ISCO}=\frac{c^3}{6^{3/2}\pi GM},
\end{equation}

causing more massive binaries to terminate their inspiral at progressively lower frequencies. Since the inspiral begins at a fixed initial frequency of 20 Hz, increasing the total mass reduces the available frequency interval over which the binary evolves before reaching the ISCO. Consequently, the duration of the inspiral decreases significantly with increasing total mass.

The numerical diagnostics confirm this interpretation. While the inspiral duration decreases from approximately 5.95 s for a $20\,M_{\odot}$ binary to only 0.03 s for a $200\,M_{\odot}$ binary, the cumulative radiated energy increases only up to intermediate masses before declining. The resulting inspiral memory therefore reaches a maximum when the enhanced gravitational-wave emission is optimally balanced by the progressively shorter inspiral duration.

The existence of this optimum demonstrates that the observable inspiral memory is governed by a balance between gravitational-wave luminosity and the duration of the inspiral within the detector's observing band. Consequently, the maximum memory is not produced by the most massive binary but by an intermediate-mass system whose inspiral remains sufficiently long while still radiating efficiently.

The diagnostic analysis presented in Fig.~\ref{fig:mass_scaling} further illustrates the origin of this behaviour. As the total mass increases, the ISCO frequency decreases monotonically, while the inspiral duration rapidly shortens. The cumulative radiated energy follows the same qualitative trend as the memory, reaching its maximum at intermediate masses before declining. These results demonstrate that the observed peak in the memory strain originates from the combined influence of the available inspiral duration and the efficiency of gravitational-wave emission. \cite{Blanchet2014,Maggiore2007,Favata2011}3

\subsection{Detector Low-Frequency Cutoff and Optimal Binary Mass}

To investigate the combined influence of detector sensitivity and binary mass, the total-mass study was repeated for several initial gravitational-wave frequencies ranging from 10 Hz to 40 Hz. Since the initial frequency represents the detector's low-frequency cutoff, this analysis directly explores how the observable inspiral memory depends on detector bandwidth.

\begin{figure}
    \centering
    \includegraphics[width=0.9\linewidth]{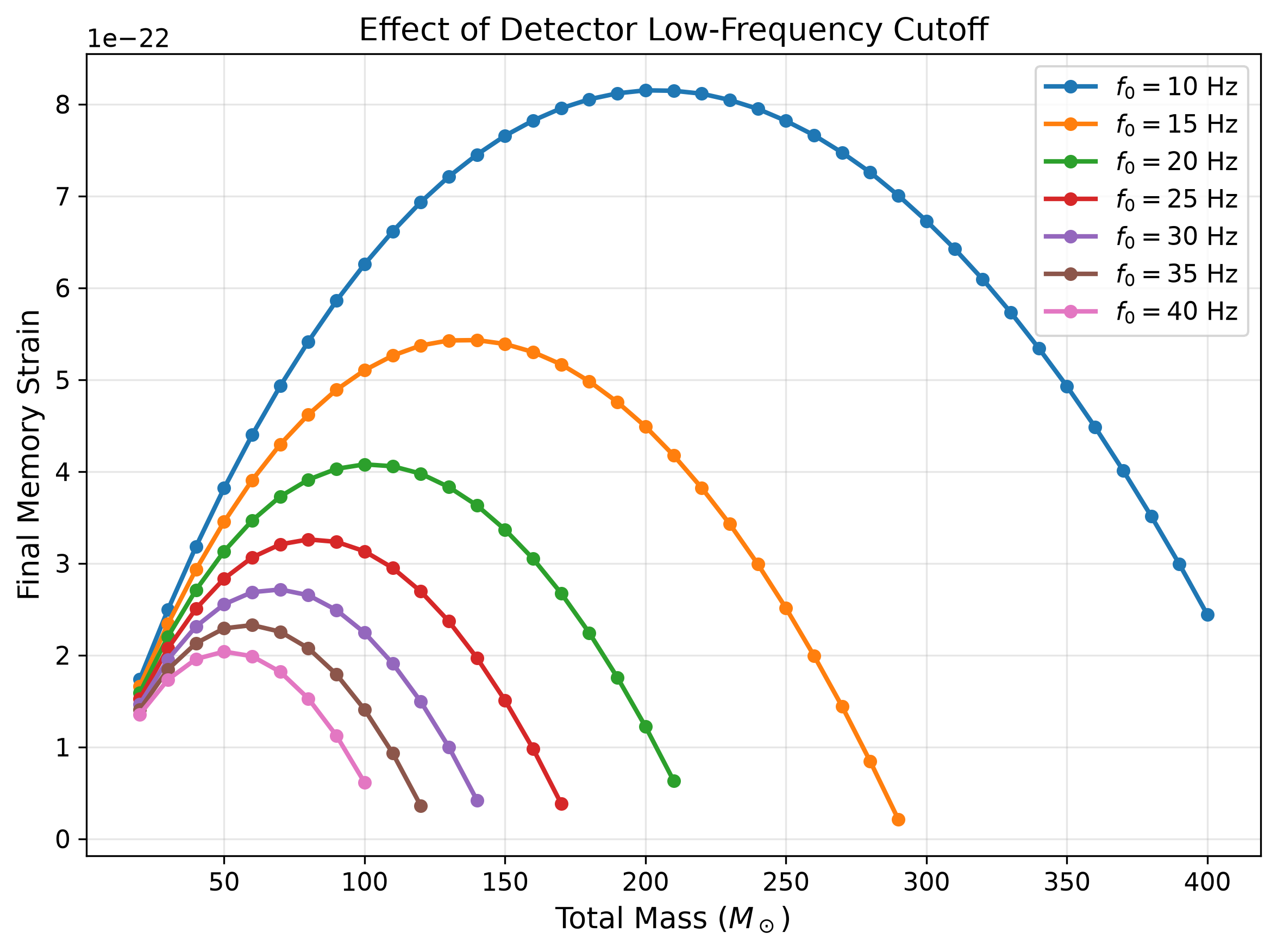}
    \caption{Accumulated Christodoulou memory strain as a function of total mass (M) across various detector low-frequency cutoffs ($f_0 \in [10, 40]\text{ Hz}$). Each curve demonstrates a distinct peak mass ($M_{\text{peak}}$) where observable inspiral memory is maximized. Lowering $f_0 $ systematically shifts $M_{\text{peak}}$ to higher total masses, following the inverse scaling relation $M_{\text{peak}} \propto f_0^{-1}$.}
    \label{fig:detector_cutoff}
\end{figure}

Figure~\ref{fig:detector_cutoff} shows that each detector cutoff exhibits a distinct optimum binary mass at which the accumulated inspiral memory reaches its maximum value. Lower detector cutoffs shift the peak toward progressively larger binary masses, whereas higher cutoffs favor lower-mass systems.

For an initial frequency of 10 Hz, the memory continues to increase throughout the investigated mass range and reaches its maximum value of approximately $8.15\times10^{-22}$ near $200\,M_{\odot}$. Increasing the detector cutoff systematically shifts the peak toward lower masses, with optimal masses of approximately $135\,M_{\odot}$, $100\,M_{\odot}$, $80\,M_{\odot}$, $65\,M_{\odot}$, $60\,M_{\odot}$, and $50\,M_{\odot}$ for detector cutoffs of 15 Hz, 20 Hz, 25 Hz, 30 Hz, 35 Hz, and 40 Hz, respectively.

The observed behaviour can be understood from the relationship between the detector bandwidth and the Schwarzschild ISCO frequency. For a fixed detector cutoff, the inspiral contributes to the observable memory only while the gravitational-wave frequency evolves from the detector's low-frequency limit to the ISCO frequency. As the binary mass increases, the ISCO frequency decreases according to

\begin{equation}
f_{\rm ISCO}\propto \frac{1}{M}.
\end{equation}

Consequently, the observable inspiral becomes progressively shorter for higher-mass systems. Lower detector cutoffs recover a larger fraction of the inspiral evolution, allowing more massive binaries to accumulate substantial nonlinear memory before reaching the ISCO. In contrast, higher detector cutoffs exclude much of the early inspiral, shifting the maximum observable memory toward lower-mass binaries whose inspiral remains within the detector bandwidth.

The optimal binary masses obtained from the numerical simulations were subsequently fitted as a function of the detector cutoff frequency. The resulting empirical relation is

\begin{equation}
    M_{\rm peak}=\frac{2006.5}{f_{\rm low}}\,M_{\odot},
    \label{eq:peakmass}
\end{equation}

where $f_{\rm low}$ is expressed in Hz. The fit reproduces the numerical results with an excellent coefficient of determination,

\begin{equation}
    R^2=0.999243,
\end{equation}

indicating that the inverse dependence provides an excellent description of the simulated data over the investigated parameter range.

Equation~(\ref{eq:peakmass}) provides a simple quantitative relationship connecting detector low-frequency sensitivity with the binary mass that maximizes the observable inspiral Christodoulou memory. To the best of our knowledge, such an empirical scaling has not been explicitly reported for inspiral-only nonlinear memory studies. Although the numerical coefficient is specific to the assumptions adopted in the present framework, the inverse dependence reflects the fundamental competition between detector bandwidth and the inspiral termination frequency.

\section{Conclusion}

In this work, we developed \texttt{GWMemoryLab}, a modular numerical framework for investigating the inspiral Christodoulou gravitational-wave memory from binary black hole systems within the post-Newtonian approximation. The framework combines binary system modeling, post-Newtonian inspiral evolution, oscillatory waveform generation, gravitational-wave energy flux calculations, and nonlinear memory estimation into a unified computational pipeline. Numerical validation confirmed the physical consistency and convergence of the implementation, demonstrating its suitability for systematic parameter-space investigations.

Using this framework, we examined the dependence of the accumulated inspiral memory on the binary mass ratio, total mass, and detector low-frequency cutoff. The simulations show that the memory increases with mass ratio and reaches its largest values for nearly equal-mass binaries, consistent with the stronger gravitational-wave emission from symmetric systems. The dependence on the initial frequency demonstrates that lower detector cutoffs recover a larger fraction of the inspiral evolution, resulting in significantly larger accumulated memory.

A particularly interesting result is the non-monotonic dependence of the inspiral memory on the total binary mass. Although more massive binaries emit stronger gravitational radiation, their inspiral terminates at lower ISCO frequencies and therefore remains within the detector bandwidth for a shorter duration. The competition between these effects produces an optimum total mass that maximizes the observable inspiral memory.

Extending this analysis to different detector low-frequency cutoffs revealed that the optimum binary mass depends systematically on the detector bandwidth. The numerical results are accurately described by the empirical scaling relation

\begin{equation}
    M_{\mathrm{peak}} = \frac{2006.5}{f_{\mathrm{low}}}\,M_{\odot},
\end{equation}

where $f_{\mathrm{low}}$ is the detector low-frequency cutoff expressed in Hz. This relation quantitatively connects detector sensitivity with the binary mass that maximizes the observable inspiral Christodoulou memory within the assumptions of the present model and provides a simple predictive tool for future parameter-space studies.

The present investigation is restricted to the inspiral phase of non-spinning, quasi-circular binary black holes described within the leading-order post-Newtonian approximation. Future developments of \texttt{GWMemoryLab} will extend the framework to include higher-order post-Newtonian corrections, spin effects, orbital eccentricity, and the merger--ringdown phases through hybrid or numerical-relativity waveforms. Such extensions will enable more realistic predictions for current and next-generation gravitational-wave observatories, including Cosmic Explorer, Einstein Telescope, and LISA.

Overall, this work demonstrates that \texttt{GWMemoryLab} provides an efficient and physically consistent platform for studying nonlinear gravitational-wave memory. Beyond validating the numerical framework, the present study establishes a quantitative connection between detector sensitivity and the binary systems that produce the strongest observable inspiral memory, offering new insight into the interplay between binary dynamics and gravitational-wave detector bandwidth.
This work demonstrates that detector characteristics are as important as binary properties in determining the observable inspiral gravitational-wave memory.

\section*{Acknowledgments}

The author acknowledges the use of OpenAI's ChatGPT as an interactive computational assistant during the development of this work. The tool was used for code development, debugging and language polishing. All scientific ideas, numerical implementations, simulations, analyses, interpretations, and final editorial decisions were independently verified and approved by the author.
The auther would like to extend gratitude to Mr. Jude Robins, Mr. Dijo Joseph  and Mr. Merilin Deep for the mental support they provided and the Government College Kattappana, Kerala, India, for providing necessary Computation facilities.

\section{Data and Code Availability}

The numerical framework developed in this work, \texttt{GWMemoryLab}, is publicly available through GitHub at

\url{https://github.com/jiswin-varghese/GWMemoryLab}

A permanent archived version of the software accompanying this manuscript is available through Zenodo:

\url{https://doi.org/10.5281/zenodo.21783195}

The repository contains the complete source code, documentation, and example scripts required to reproduce the numerical results presented in this work.


\end{document}